\documentclass[smallextended]{svjour3}       

\usepackage{amsfonts}
\usepackage{amssymb}
\usepackage{amsmath}
\usepackage{hyperref}
\usepackage{graphicx,subfigure}
\usepackage{latexsym}
\usepackage{mathptmx}      

\journalname{Quantum Information Processing}

\newcommand{\bldF}{\mathbf{F}}
\newcommand{\bldx}{\mathbf{x}}
\newcommand{\bldy}{\mathbf{y}}
\newcommand{\bldz}{\mathbf{z}}
\newcommand{\bldG}{\mathbf{G}}

\newcommand{\Dbeta}{\Delta\boldsymbol\beta}
\newcommand{\Dlambda}{\Delta\boldsymbol{\lambda}}

\newcommand{\DbfF}{\Delta \mathbf{F}}

\newcommand{\Dbfy}{\Delta \mathbf{y}}

\newcommand{\sz}{\sigma_z}
\newcommand{\sx}{\sigma_x}
\newcommand{\sy}{\sigma_y}

\newcommand{\bsig}{\boldsymbol\sigma}

\newcommand{\bldcdot}{\mbox{\Large\boldmath $\cdot$}}

\newcommand{\calG}{\mathcal{G}}
\newcommand{\calH}{\mathcal{H}}
\newcommand{\calO}{\mathcal{O}}
\newcommand{\calF}{\mathcal{F}}

\newenvironment{coi}{\noindent}{}

\begin{document}

\title{High-fidelity quantum state preparation using neighboring optimal 
control}

\author{Yuchen Peng, Frank Gaitan}

\institute{Frank Gaitan \at
                Laboratory for Physical Sciences, 8050 Greenmead Dr, 
                College Park, MD 20740\\
                Tel.: 301-935-6531\\
                Fax: 301-935-6723\\
                \email{fgaitan@lps.umd.edu}\\
               ORCID ID: 0000-0002-2537-803X
                  \and
                Yuchen Peng \at
                Department of Physics, University of Maryland, College Park,
                MD 20742
                }

\date{\today}

\maketitle

\begin{abstract}
We present an approach to single-shot high-fidelity preparation of an $n$-qubit
state based on neighboring optimal control theory. This represents a new 
application of the neighboring optimal control formalism which was originally 
developed to produce single-shot high-fidelity quantum gates. To illustrate 
the approach, and to provide a proof-of-principle, we use it to prepare 
the two qubit Bell state $|\beta_{01}\rangle = (1/\sqrt{2})\left[\, |01\rangle 
+ |10\rangle\,\right]$ with an error probability $\epsilon\sim 10^{-6}$ 
($10^{-5}$) for ideal (non-ideal) control. Using standard methods in the 
literature, these high-fidelity Bell states can be leveraged to fault-tolerantly
prepare the logical state $|\overline{\beta}_{01}\rangle$.

\keywords{Quantum state preparation, quantum optimal control theory, 
                   neighboring optimal control, Bell states}

\PACS{03.65.Ud, 03.67.Ac, 03.67.Bg, 03.67.Hk}
\subclass{81P68, 81Q93, 93C15}
\end{abstract}

\section{Introduction\label{sec1}}

In optimal control theory the problem is to determine a control field profile
$\bldF_{\ast}(t)$ that optimizes system performance relative to a set of
design criteria. A cost function is introduced that quantifies the degree to
which a particular assignment of the control and system variables satisfy
these criteria, with optimal assignment being one of minimum cost \cite{Sten}.
In the quantum version of this problem the optimal control $\bldF_{\ast}(t)$
drives an optimal unitary transformation $U_{\ast}$ that yields a 
high-fidelity approximation to a target unitary $U_{tgt}$. The target $U_{tgt}$ 
might represent a desired quantum gate, or it might be used to prepare a quantum
state $|\psi_{tgt}\rangle = U_{tgt}|\psi_{in}\rangle$, given an easily prepared 
initial state $|\psi_{in}\rangle$. Note that a perturbation of the quantum 
dynamics can cause the optimal control $\bldF_{\ast}(t)$ to become nonoptimal.
If the perturbation is small, however, the optimal control problem can be
linearized about the unperturbed optimal control, and a family of perturbed
optimal controls found from a single feedback control law. In the classical 
literature this perturbed problem is referred to as neighboring optimal control 
(NOC) \cite{Sten}.

In this paper we present a general approach to single-shot high-fidelity quantum
state preparation based on NOC theory. This represents a new application of 
the neighboring optimal control formalism which was originally developed to 
produce single-shot high-fidelity quantum gates. In Section~\ref{sec2} we 
formulate the NOC problem and determine the Euler-Lagrange equations whose 
solution determines the optimal control $\bldF_{\ast}(t)$ and the resulting 
optimal quantum state $|\psi_{\ast}\rangle$. In Section~\ref{sec3} (\ref{sec4}) 
we illustrate the general approach by using it to prepare a high-fidelity 
approximation to the two qubit Bell state $|\beta_{01}\rangle = (1/\sqrt{2})
\left[\, |01\rangle + |10\rangle\,\right]$ with error probability $\epsilon \sim 
10^{-6}$ ($10^{-5}$), assuming ideal (non-ideal) control. The high-fidelity of
the final state provides proof-of-principle for the performance gains 
possible using NOC, even in the presence of imperfect control. Note that 
standard methods in the literature \cite{knill} can leverage these high-fidelity
Bell states to fault-tolerantly prepare the logical state 
$|\overline{\beta}_{01}\rangle$. We close in Section~\ref{sec5} with a 
discussion of our results. Finally, in Appendix~\ref{AppA}, we briefly review 
the essential features of a type of non-adiabatic rapid passage used in 
Sections~\ref{sec3} and \ref{sec4}, and specify the Hamiltonian used to 
drive the two-qubit dynamics.

\section{High-fidelity state preparation using NOC}
\label{sec2}

In this Section we present an approach to $n$-qubit state preparation based on
NOC. The task is to find a control field $\bldF (t)$ that produces a quantum 
state $|\psi_{f}\rangle$ which is a high-fidelity approximation to a target 
state $|\psi_{tgt}\rangle$. In Section~\ref{sec2_1} we transform the state 
preparation problem to that of implementing a high-fidelity approximation 
$U_{f}$ to a desired unitary gate $U_{tgt}$. Sections~\ref{sec2_2} and
\ref{sec2_3} summarize earlier work \cite{Peng&Gaitan} that used 
NOC theory to produce such a high-fidelity approximation. These 
Sections formulate the optimization problem whose solution yields the optimal 
control field $\bldF_{\ast}(t)$ and unitary gate $U_{\ast}$; derive the 
equations governing the optimization; and describe a procedure for 
obtaining a solution. 

\subsection{Reformulating the control problem}
\label{sec2_1}

Consider an $n$-qubit system whose Hamiltonian is a functional of a control 
field $\bldF (t)$:
\begin{displaymath}
\calH (t) = \calH [\bldF (t)].
\end{displaymath}
Throughout this paper we assume: (i)~the control $\bldF (t)$ acts for times
$-T/2\leq t\leq T/2$; and (ii)~the control duration $T$ is much shorter than the
longitudinal ($T_{1}$) and transverse ($T_{2}$) relaxation times so that a qubit
is weakly decohering and a state vector description is appropriate. See 
Section~\ref{sec5} for further discussion of assumption~(ii).

Suppose the initial $n$-qubit state is the ``all-zeros'' computational basis 
state (CBS) $|\psi (-T/2)\rangle = |0\cdots 0\rangle$. The control field 
$\bldF (t)$ drives a unitary transformation $U(t)$ that produces the final 
state
\begin{equation}
|\psi_{f}\rangle = U_{f}|0\cdots 0\rangle,
\label{actualState}
\end{equation}
where $U_{f}\equiv U(T/2)$. Let $\{\, |\beta_{i}\rangle : i = 1, \ldots , 
M = 2^{n}\}$ be an orthonormal basis for the $n$-qubit Hilbert space
that has $|\beta_{1}\rangle = |\psi_{tgt}\rangle$. Defining
the target gate $U_{tgt}$ as the unitary matrix whose columns are the basis 
vectors $\{ |\beta_{i}\rangle\}$,
\begin{equation}
U_{tgt} = \left(
                         \begin{array}{ccccc}
                            \vline\vspace{0.6 em} & & \vline & & \vline\\
                             |\beta_{1}\rangle\vspace{0.4 em} & \cdots & 
                                   |\beta_{i}\rangle & \cdots & |\beta_{M}
                                         \rangle \\
                            \vline & & \vline & & \vline
                        \end{array}
                 \right),
\label{targetGate}
\end{equation}
we see that, by construction,
\begin{equation}
 |\psi_{tgt}\rangle =  U_{tgt} |0\cdots 0\rangle . 
\label{targetState}
\end{equation}
These definitions reformulate the problem of preparing a high-fidelity
approximation $|\psi_{f}\rangle$ to the target state $|\psi_{tgt}\rangle$ to 
that of implementing a high-fidelity approximation $U_{f}$ to the target gate 
$U_{tgt}$ (Eq.~(\ref{targetGate})). A solution to this problem was found in 
Ref.~\cite{Peng&Gaitan} using NOC. In the remainder of this Section we 
summarize that solution and refer the reader to Ref.~\cite{Peng&Gaitan} for a 
detailed presentation.

\subsection{Quantum dynamics of NOC}
\label{sec2_2}

Ref.~\cite{Peng&Gaitan} showed how NOC could be used to improve the performance
of a good quantum gate. The starting point is a nominal control $\bldF_{0}(t)$ 
which enacts a unitary $U_{0}(t)$ such that $U_{0,f} = U_{0}(T/2)$ is a good
approximation to the target $U_{tgt}$. Then $U^{\dagger}_{0,f}$ is close to
the inverse of $U_{tgt}$ so that
\begin{equation}
U^{\dagger}_{0,f}U_{tgt} = I -i\delta\beta + \calO (\Delta^{2}).
\label{nearInverse}
\end{equation}
NOC was then used to determine a control modification $\Delta\bldF (t)$ which
yields a new control $\bldF (t) = \bldF_{0}(t) + \Delta\bldF (t)$ which drives a
unitary $U(t)$ that provides a better approximation $U_{f} = U(T/2)$ to 
$U_{tgt}$. Since $\bldF_{0}(t)$ is a good control, $\Delta\bldF (t)$ is expected
to be small. To begin, we write $U(t)$ as
\begin{equation}
U(t) = U_{0}(t)\delta U(t) ,
\label{Udef}
\end{equation}
where
\begin{equation}
\delta U(t) = I -i\delta A(t) + \calO (\Delta^{2}).
\label{delUdef}
\end{equation}
Substituting Eqs.~(\ref{Udef}) and (\ref{delUdef}) into the Schrodinger equation
($\hbar = 1$)
\begin{equation}
i\frac{d}{dt} U(t) = \calH [\bldF (t)] U(t),
\label{SchEq}
\end{equation}
gives the equation of motion for $\delta A(t)$,
\begin{equation}
\frac{d}{dt}\delta A(t) = \sum_{j=1}^{3} \overline{G}_{j}(t)\Delta F_{j}(t)
                                              + \calO (\Delta^{2}) ,
\label{AEoM}
\end{equation}
with
\begin{equation}
\overline{G}_{j}(t) = U^{\dagger}_{0}(t)\calG_{j}(t)U_{0}(t)
\end{equation}
and
\begin{equation}
\calG_{j}(t) = \left. \frac{\delta H}{\delta F_{j}}\right|_{\bldF_{0}(t)} .
\end{equation}

It proves convenient to vectorize all matrices by concatenating their columns.
For example, let $\Delta\bldx (t)$ be the vectorization of $\delta A(t)$:
\begin{displaymath}
\Delta\bldx (t) = \left(
                                   \begin{array}{c}
                                        \delta A_{\:\, \bldcdot ,\; 1}\vspace{0.5 em}(t)\\
                                         \vdots\vspace{0.5 em} \\
                                        \delta A_{\:\, \bldcdot ,\; M}\vspace{0.5 em}(t)\\
                                   \end{array}
                           \right) .
\end{displaymath}
Similarly, we write $\delta\beta$ and $\overline{G}_{j}(t)$ as
the vectors $\Dbeta$ and $\bldG_{j}(t)$. Lastly, we define the $2^{n}
\times 3$ matrix $G(t)$ as
\begin{displaymath}
G(t) = \left(
                   \begin{array}{ccc}
                        \vline\vspace{0.5 em} & \vline & \vline \\
                        \bldG_{1}(t)\vspace{0.5 em} & \bldG_{2}(t) & 
                                                    \bldG_{3}(t) \\
                        \vline & \vline & \vline 
                   \end{array}
           \right).
\end{displaymath}
With these definitions, and introducing $\Delta\bldy (t) = \Delta\bldx (t) - 
\Dbeta$, Eq.~(\ref{AEoM}) becomes
\begin{equation}
\frac{d}{dt}\Delta\bldy (t) = G(t)\Delta\bldF (t),
\label{vectorEoM}
\end{equation}
with initial condition  
\begin{equation}
\Delta\bldy (-T/2) = -\Dbeta ,
\end{equation}
which is a consequence of $\delta A(-T/2) = 0$.

\subsection{Optimal control problem}
\label{sec2_3}

To determine the optimal control modification $\Delta\bldF (t)$ and the 
associated improved unitary gate $U_{f}$ we introduce a cost function $J$
whose variation yields the equations governing the optimization:
\begin{eqnarray}
J & = & \Delta\bldy^{\dagger}(T/2)\Delta\bldy (T/2) \nonumber \\
   & & \hspace{0.4in} + \int_{-T/2}^{T/2} dt\left[ \Delta\bldy^{\dagger}(t)
             Q(t)\Delta\bldy (t) + \frac{1}{2}\Delta\bldF^{T}(t)R(t)
                 \Delta\bldF (t)\right] \nonumber \\
& & \hspace{0.8in} + \int_{-T/2}^{T/2} dt\left[ \Dlambda^{\dagger}(t) 
         \left\{ G(t)\DbfF (t) - \frac{d}{dt}\Delta\bldy (t)\right\} + h.c.\right] .
\end{eqnarray}
The cost function: (i)~penalizes controls which yield $U_{f}\neq U_{tgt}$ and
large $\DbfF (t)$ and $\Dbfy (t)$; and (ii)~introduces a Lagrange
multiplier $\Dlambda (t)$ that insures the optimization does not violate the 
Schrodinger dynamics. The matrices $Q(t)$ and $R(t)$ are required to be
positive-definite and Hermitian, though are otherwise at our disposal.
Requiring that $J$ be stationary under variation of $\Dbfy (t)$, $\Dbfy (T/2)$,
$\DbfF (t)$, and $\Dlambda (t)$ gives the optimization equations of motion 
(EOM):
\begin{subequations}
\label{totEOM}
\begin{eqnarray}
 & &   \DbfF (t)  =   R^{-1}(t) G^{\dagger}(t)\Dlambda (t) 
  \label{feedback}\\
\frac{d}{dt}\Dlambda (t) + Q(t)\Dbfy (t) = 0 
   & \hspace{0.25in} ; \hspace{0.25in} & \Dlambda (T/2) = 
                                  \Dbfy (T/2) \label{dellamIC}\\
\frac{d}{dt}\Dbfy (t) - G(t)\DbfF (t) = 0 
  & \hspace{0.0in} ; \hspace{0.0in} &
          \Dbfy (-T/2) = -\Dbeta . \label{delyIC}
\end{eqnarray}
\end{subequations}

To simplify the solution of Eqs.~(\ref{totEOM}) we introduce the Ricatti matrix 
$S(t)$ via
\begin{equation}
\Dlambda (t) = S(t)\Dbfy (t) .
\label{Ricattidef}
\end{equation}
Differentiating Eq.~(\ref{Ricattidef}) and using Eqs.~(\ref{totEOM}) yields the 
differential equation for $S(t)$:
\begin{equation}
\frac{dS}{dt} = -Q + SGR^{-1}G^{\dagger}S .
\label{RicattiEq}
\end{equation}
Its initial condition follows from Eq.~(\ref{Ricattidef}) evaluated at $t=T/2$
and Eq.~(\ref{dellamIC}):
\begin{equation}
S(T/2) = I.
\label{RicattiIC}
\end{equation}
Inserting Eq.~(\ref{Ricattidef}) into Eq.~(\ref{feedback}) gives the feedback 
control law
\begin{equation}
\DbfF (t) = -C(t)\Dbfy (t),
\label{feedcontrollaw}
\end{equation}
where 
\begin{equation}
C(t) = R^{-1}(t)G^{\dagger}(t)S(t) 
\label{gainMatrix}
\end{equation}
is the control gain matrix. 

Eqs.~(\ref{delyIC}), and (\ref{RicattiEq})--(\ref{gainMatrix}) form 
an equivalent set of EOM which are easier to solve. The first step is to solve 
Eqs.~(\ref{RicattiEq}) and (\ref{RicattiIC}) for $S(t)$. Since these Eqs.\ are 
independent of $\Dbfy (t)$ and $\DbfF (t)$, they can be solved immediately. 
Knowing $S(t)$, Eq.~(\ref{gainMatrix}) allows the gain matrix $C(t)$ to be 
determined. Substituting Eq.~(\ref{feedcontrollaw}) into Eq.~(\ref{delyIC}) 
gives
\begin{equation}
\frac{d}{dt}\Dbfy = -GC\Dbfy .
\label{delyeom}
\end{equation}
This can be integrated for $\Dbfy (t)$ subject to the initial condition
$\Dbfy (-T/2) = -\Dbeta$. Note that $\Dbeta$ is determined by 
Eq.~(\ref{nearInverse}) (for an example, see Section~\ref{sec3}). With
$\Dbfy (t)$ in hand, the control modification $\DbfF (t)$ follows from
the feedback control law (Eq.~(\ref{feedcontrollaw})). The new control is then 
 $\bldF (t) = \bldF_{0}(t) + \DbfF (t)$, and the new gate $U_{f} = U(T/2)$ is 
found by plugging $\bldF (t)$ into the Hamiltonian $\calH [\bldF (t)]$ and 
integrating the Schrodinger equation (Eq.~(\ref{SchEq})) for $U(t)$. The 
improved state is then
\begin{equation}
|\psi_{f}\rangle = U_{f}|0\cdots 0\rangle
\label{newstate}
\end{equation}
with fidelity
\begin{equation}
\calF = |\langle\psi_{f}|\psi_{tgt}\rangle |
\label{fidelity}
\end{equation}
and error probability
\begin{equation}
\epsilon = 1 - \calF^{2} .
\label{errprob}
\end{equation}
From an experimental point of view, the essential result is the new control 
$\bldF (t)$ which drives the state $|0\cdots 0\rangle$ to a high-fidelity
approximation to $|\psi_{tgt}\rangle$. We illustrate this approach
in the following Section.

\section{Bell state preparation---ideal control}
\label{sec3}

Here we show how the approach to quantum state preparation described in
Section~\ref{sec2} can be used to prepare a high-fidelity approximation to the
Bell state 
\begin{displaymath}
|\beta_{01}\rangle = (1/\sqrt{2})\left[ |01\rangle + |10\rangle
\right]. 
\end{displaymath}
As that discussion was general, no specific 
form was assumed for the Hamlitonian $\calH [\bldF (t)]$. In Sections~\ref{sec3}
and \ref{sec4} we assume the two-qubit Hamiltonian $\calH_{2}$ couples each
qubit to the control field $\bldF (t)$ through the Zeeman interaction, and 
couples the qubits through an anisotropic Heisenberg interaction. In the lab 
frame ($\hbar = 1$):
\begin{equation}
\calH_{2}[\bldF (t)] = -\sum_{i=1}^{3}\frac{\gamma_{i}}{2}\bsig^{i}\cdot
                                     \bldF (t)  -\frac{\pi}{2}\left[ 
                                       J_{z}\sz^{1}\sz^{2} + J_{xy}\left( 
                                         \sx^{1}\sx^{2} + \sy^{1}\sy^{2}\right)
                                         \right] .
\label{labHam_v0}
\end{equation}
Since $\bldF (t) = \bldF_{0}(t) + \DbfF (t)$, we have
\begin{equation}
\calH_{2}[\bldF (t)] = \calH^{0}_{2}[\bldF_{0}(t)] + \sum_{j=1}^{3}
                                      \calG_{j}(t)\Delta F_{j}(t) ,
\label{labHam_v1}
\end{equation}
where
\begin{equation}
\calH_{2}^{0}[\bldF_{0} (t)] = -\sum_{i=1}^{3}\frac{\gamma_{i}}{2}\bsig^{i}\cdot
                                     \bldF_{0}(t)  -\frac{\pi}{2}\left[ 
                                       J_{z}\sz^{1}\sz^{2} + J_{xy}\left( 
                                         \sx^{1}\sx^{2} + \sy^{1}\sy^{2}\right)
                                         \right] ,
\label{nomHam}
\end{equation}
and
\begin{equation}
\calG_{j}(t) = -\frac{1}{2}\sum_{i=1}^{2}\gamma_{i}\sigma^{i}_{j} .
\label{Gjdef}
\end{equation}
Thus $\calH^{0}_{2}[\bldF_{0}(t)]$ is the nominal Hamiltonian that drives
the good starting gate $U_{0}(t)$ introduced in Section~\ref{sec2_2} and the
dynamical contribution of the control modification $\DbfF (t)$ is contained in 
the second term on the RHS of Eq.~(\ref{labHam_v1}).

\subsection{Nominal dynamics}
\label{sec3_1}

As noted earlier, the nominal control $\bldF_{0}(t)$ enacts a unitary 
transformation $U_{0}(t)$ which maps the initial state $|\psi_{0}(-T/2)\rangle
 = |00\rangle$ to the final state $|\psi_{0}(T/2)\rangle = |\psi_{0,f}\rangle$:
\begin{displaymath}
|\psi_{0,f}\rangle = U_{0,f} |00\rangle,
\end{displaymath}
where $|\psi_{0,f}\rangle$ is a good approximation to the Bell state
$|\beta_{01}\rangle$. In this subsection we use a form of non-adiabatic rapid 
passage known as twisted rapid passage (TRP) \cite{orgTRP,expTRP} to provide
the nominal control $\bldF_{0}(t)$. The reader will find a summary of TRP 
essentials in Appendix~\ref{AppA}. We stress that NOC theory is not
restricted to this particular type of starting control; any other control can 
be used so long as it produces a sufficiently good approximation to 
$|\beta_{01}\rangle$. See Section~\ref{sec5} for further discussion of this
point.

In the lab frame the TRP control is
\begin{equation}
\bldF_{0}(t) = B_{0}\hat{\bldz} + B_{rf}\cos\phi_{rf}(t)\hat{\bldx}
                        -B_{rf}\sin\phi_{rf}(t)\hat{\bldy}.
\label{TRPcontroldef}
\end{equation}
The transformation to the detector frame \cite{expTRP,quadTRP} is carried out 
by the unitary
\begin{displaymath}
U_{det}(t) = \exp\left[\frac{i}{2}\phi_{det}(t)\left( \sz^{1} +\sz^{2}\right)
                            \right] .
\end{displaymath}
Appendix~\ref{AppA} (see also Ref.~\cite{orgTRP}) shows that $\phi_{rf}(t) = 
\phi_{det}(t) -\phi_{trp}(t)$, where $\dot{\phi}_{trp}(t)$ is the instantaneous 
rate at which the control field $\bldF_{0}(t)$ rotates about the $z$-axis in the
detector frame (see Eq.~(\ref{1qbtHam}) below). We restrict ourselves to 
quartic twist $\phi_{trp}(t) = (1/2)Bt^{4}\equiv \phi_{4}(t)$ in this paper. 
The nominal Hamiltonian in the detector frame is then
\begin{equation}
\calH_{det}^{0}(t) = U^{\dagger}_{det}(t)\calH^{0}_{2}(t) U_{det}(t) 
                 - iU^{\dagger}_{det}(t)\frac{d}{dt}U_{det}(t) .
\label{detHam}
\end{equation}

It proves useful for the numerical simulations to recast the Schrodinger
dynamics into dimensionless form. This is done in Appendix~\ref{AppA} (see also
Ref.~\cite{Peng&Gaitan}) with the result
\begin{equation}
\overline{\calH}^{0}_{det}(\tau ) = \calH_{1}^{0}(\tau ) + \calH^{0}_{int}
           (\tau ),
\label{dimHam}
\end{equation}
where the one-qubit term is ($\phi_{4}(\tau )=(\eta_{4}/2\lambda )\tau^{4}$)
\begin{eqnarray}
\calH_{1}^{0}(\tau ) & = & \left[ -\frac{(d_{1}+d_{2})}{2} + 
                  \frac{\tau}{\lambda}\right]\sz^{1} -\frac{d_{3}}{\lambda}\left[
                     \cos\phi_{4}(\tau )\sx^{1} + \sin\phi_{4}(\tau )\sy^{1}
                       \right] \nonumber \\
 & & \hspace{0.5in}+\left[ -\frac{d_{2}}{2} + 
                  \frac{\tau}{\lambda}\right]\sz^{2} -\frac{1}{\lambda}\left[
                     \cos\phi_{4}(\tau )\sx^{2} + \sin\phi_{4}(\tau )\sy^{2}
                       \right]
\label{1qbtHam}
\end{eqnarray}
and the interaction term is 
\begin{equation}
\calH_{int}^{0}(\tau ) = -\frac{\pi}{2}\left[ d_{z}\sz^{1}\sz^{2} + d_{xy}\left(
                              \sx^{1}\sx^{2} + \sy^{1}\sy^{2}\right)\right] .
\label{intHam}
\end{equation}
The parameters $d_{1},d_{2},d_{3},d_{z}, d_{xy}$ are dimensionless 
versions of the qubit couplings to the TRP control as well as to each other and
are defined below Eq.~(\ref{AppdimHam}). The remaining parameters $\tau$ 
(dimensionless time), $\lambda$ (dimensionless TRP inversion rate), $\eta_{4}$ 
(dimensionless twist strength), and $\tau_{0}$ (dimensionless TRP inversion 
time) are defined above Eq.~(\ref{AppdimHam}) and they determine the actual 
TRP control profile $\bldF_{0}(t)$ applied to the qubits.

For the TRP dynamics to produce a good approximation to $|\beta_{01}\rangle$
we must search for suitable values for the parameters appearing in 
$\overline{\calH}^{0}_{det}(\tau )$. We used simulated annealing to find a
parameter assignment that minimized the state preparation error probability
\begin{equation}
\epsilon_{0} = 1 - |\langle\psi_{0,f}|\beta_{01}\rangle|^{2} .
\label{errprobnom}
\end{equation}
As in the two-qubit simulations done in Ref.~\cite{Peng&Gaitan}, we set 
$\tau_{0}=120$ in this paper. Table~\ref{table1}
\begin{table}
\caption{\label{table1}Simulated annealing (SA) was used to find values for the
parameters $\eta_{4},\lambda , d_{1},d_{2},d_{3},d_{z},d_{xy}$ that
minimize the error probability for preparing a high-fidelity approximation to 
the Bell state $|\beta_{01}\rangle$. MATLAB was used to implement SA with an 
initial (dimensionless) temperature $T_{0} = 100$ and an exponential annealing 
schedule with a reduction factor of $0.95$. The starting
parameter values were $\eta_{4} = 10^{-4}$, $\lambda = d_{1} = d_{2} =
d_{3} = d_{z} = d_{xy} = 1$. The parameter values found are listed below and 
produced a state preparation error probability $\epsilon_{0} = 6.68\times 
10^{-4}$.}
\begin{center}
\begin{tabular}{|c|c|c|c|c|c|c|} \hline
$\eta_{4}$ & $\lambda$  & $d_{1}$ & $d_{2}$ & $d_{3}$  
                  & $d_{z}$ & $d_{xy}$\\ \hline
$4.526\times 10^{-4}$ & $9.579$ & $1.386$ & $9.622$ & $8.905$ & $0.918$
            & $4.331$ \\\hline
\end{tabular}
\end{center}
\end{table}
lists the minimizing control parameter values found. Numerical integration of 
the two-qubit Schrodinger equation using the TRP control Hamiltonian
$\overline{\calH}^{0}_{det}(\tau )$ gave the final state $|\psi_{0,f}
\rangle$ in the detector frame which was then transformed back to the lab frame.
The state obtained (in the computational basis) is
\begin{equation}
|\psi_{0,f}\rangle = \left(
                                               \begin{array}{c}
                                                     -0.0070 - 0.0066\, i\\
                                                     -0.2503 - 0.6870\, i\\
                                                     -0.2006 - 0.667\, i\\
                                                      0.0080 + 0.0164\, i
                                              \end{array}
                                      \right)
   = -e^{1.28i}\sqrt{1-\epsilon_{0}}|\beta_{01}\rangle 
             + \calO (\sqrt{\epsilon_{0}}),
\label{TRPprepstate}
\end{equation}
with (state preparation) error probability 
\begin{equation}
\epsilon_{0} = 6.68\times 10^{-4}
\label{actTRPerrprob}
\end{equation}
and associated fidelity 
\begin{equation}
\calF_{0} = \sqrt{1-\epsilon_{0}} = 0.9997 .
\label{TRPfidelity}
\end{equation}
The TRP nominal control thus provides an excellent starting point for the
NOC approach to preparing $|\beta_{01}\rangle$.

\subsection{NOC dynamics}
\label{sec3_2}

The TRP driven dynamics with parameter values as in Table~\ref{table1} map
the initial state $|00\rangle$ to $U_{0,f}|00\rangle = |\psi_{0,f}\rangle$ 
(see Eq.~(\ref{TRPprepstate})):
\begin{equation}
U_{0,f}|00\rangle = -e^{1.28i}\,\sqrt{1-\epsilon_{0}}\, |\beta_{01}\rangle
                  +\calO (\sqrt{\epsilon_{0}}).
\label{TRPprepstate_v1}
\end{equation}
The same TRP control drives the remaining CBS $|ij\rangle$ to
\begin{subequations}
\label{remainingBellstates}
\begin{eqnarray}
U_{0,f}|01\rangle &  = \left(
                                          \begin{array}{c}
                                             -0.7054 -0.1678i\\
                                              0.0036 + 0.0139i\\
                                            -0.0130 - 0.0143i\\
                                           -0.6694 -0.1595i
                                         \end{array}
                                 \right)
                   	        & =  -e^{0.24i}\sqrt{1-\epsilon_{0}^{00}}
                  |\beta_{00}\rangle + \calO (\sqrt{\epsilon_{0}^{00}})  \\
U_{0,f}|10\rangle &  = \left(
                                          \begin{array}{c}
                                             -0.6686 -0.1591i\\
                                              0.0188 + 0.0311i\\
                                              0.0051 - 0.0182i\\
                                              0.7053 +0.1683i
                                         \end{array}
                                 \right)
                   	        & =  -e^{0.24i}\sqrt{1-\epsilon_{0}^{10}}
                  |\beta_{10}\rangle + \calO (\sqrt{\epsilon_{0}^{10}})  \\
U_{0,f}|11\rangle &  = \left(
                                          \begin{array}{c}
                                              0.0374 -0.0172i\\
                                              0.1214 + 0.6853i\\
                                             -0.1267 - 0.7055i\\
                                             -0.0115 +0.0006i
                                         \end{array}
                                 \right)
                   	        & =  e^{1.39i}\sqrt{1-\epsilon_{0}^{11}}
                  |\beta_{11}\rangle + \calO (\sqrt{\epsilon_{0}^{11}}), 
\end{eqnarray}
\end{subequations}
where $\epsilon_{0}^{ij}\sim\epsilon_{0}$.
For the reader's convenience we list all two-qubit Bell states in 
Table~\ref{table2}.
\begin{table}
\caption{\label{table2}Two-qubit Bell states $|\beta_{ij}\rangle$.}
\begin{center}
\begin{tabular}{|c|} \hline
$|\beta_{00}\rangle = \left[ |00\rangle + |11\rangle\right]/\sqrt{2}$\\
$|\beta_{10}\rangle = \left[ |00\rangle - |11\rangle\right]/\sqrt{2}$\\
$|\beta_{01}\rangle = \left[ |01\rangle + |10\rangle\right]/\sqrt{2}$\\
$|\beta_{11}\rangle = \left[ |01\rangle - |10\rangle\right]/\sqrt{2}$  \\\hline
\end{tabular}
\end{center}
\end{table}
We see that TRP provides excellent approximations for all four Bell states.
Based on Eqs.~(\ref{TRPprepstate_v1}) and (\ref{remainingBellstates}) we use 
the following target unitary for the NOC state preparation procedure
\begin{equation}
U_{tgt} = \left(
                         \begin{array}{cccccccc}
                              \vline\vspace{0.6 em} & & \vline & & \vline & & \vline  & \\
                            -e^{1.28i} |\beta_{01}\rangle\vspace{0.4 em} &  & 
                                   -e^{0.24i}|\beta_{00}\rangle &  & -e^{0.24i}|\beta_{10}
                                         \rangle & & e^{1.39i}|\beta_{11}\rangle &  \\
                            \vline & & \vline & & \vline & & \vline & 
                        \end{array}
                \right) .
\end{equation}
As discussed in Section~\ref{sec2_3}, to obtain the control modification 
$\DbfF (t)$ we first integrate the Ricatti equation (Eqs.~(\ref{RicattiEq}) 
and (\ref{RicattiIC})). To that end, we chose $R(\tau ) =
rG^{\dagger}(\tau ) G(\tau )$, $Q(\tau ) = G(\tau )\left( G^{\dagger}(\tau ) 
G(\tau )\right)^{-1}G^{\dagger}(\tau )/r$, and $r=70$. The solution to the
Ricatti equation is then $S(\tau ) = I_{16\times 16}$, the $16\times 16$ 
identity matrix. Eq.~(\ref{gainMatrix}) then gives the feedback control matrix
$C(\tau )$ which allows Eq.~(\ref{delyeom}) to be integrated subject to the
initial condition $\Dbfy (-T/2) = -\Dbeta$. The constant vector $\Dbeta$ is 
found by concatenating the columns of $\delta\beta$ (see 
Eq.~(\ref{nearInverse})). With $\Dbfy (\tau )$ in hand, 
Eq.~(\ref{feedcontrollaw}) determines the control modification $\DbfF (\tau )$.
The improved control $\bldF (\tau ) = \bldF_{0}(\tau ) + \DbfF (\tau )$ is then
plugged into the Schrodinger equation using Eq.~(\ref{labHam_v1}) for $\calH_{2}
[\bldF (\tau )]$. With initial state $|00\rangle$, the resulting final state is
\begin{equation}
|\psi^{NOC}_{f}\rangle = \left(
                                                   \begin{array}{c}
                                                       -0.0043 - 0.0043i\\
					  -0.2044 - 0.6849i\\
 					  -0.2006 - 0.6701i\\
 					   0.0059 + 0.0118i\\
                                                  \end{array}
                                           \right)
\label{NOCBellstate}
\end{equation}
which has error probability
\begin{equation}
\epsilon_{NOC} = 1 -|\langle\psi^{NOC}_{a}|\beta_{01}\rangle|^{2} =
           2.58\times 10^{-6}
\label{NOCerrprob}
\end{equation}
and fidelity
\begin{equation}
\calF_{NOC} = 0.999\,999.
\label{NOCfidelity}
\end{equation}
We see that NOC has significantly improved the approximation to $|\beta_{01}
\rangle$, reducing the error probability by two orders-of-magnitude
and adding three $9$'s to the fidelity.

\begin{figure}[!htb]
\centering
\includegraphics[trim=3.5cm 8.5cm 4.5cm 8.5cm,clip,
   width=.45\textwidth]{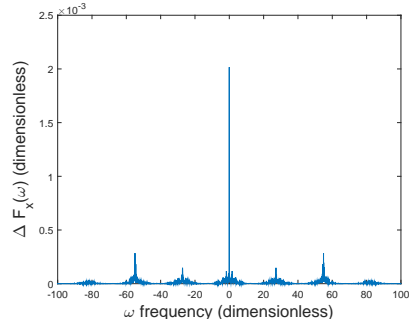}\\
\caption{The Fourier transform $\Delta \calF_{x}(\omega)$
of $\Delta F_{x}(\tau )$ used to prepare a NOC improved approximation to
the state $|\beta_{01}\rangle$. Here $\omega$ is dimensionless frequency.
\label{figure1}}
\end{figure}

We can estimate the bandwidth needed for $\DbfF (t)$ from Figure~\ref{figure1}
which shows the Fourier transform $\Delta\calF_{x}(\omega )$ of $\DbfF_{x}
(\tau )$; the $y$- and $z$-components behave comparably. We see that 
$\Delta\calF_{x}(\omega )$ is reduced to $1$-$2\%$ of its peak value for
$|\omega |\gtrsim 60$, giving a \textit{dimensionless\/} bandwidth $\Delta\omega
\sim 60$. Choosing a control operation time $T = 5\mu s$ which corresponds to
a dimensionless inversion time $\tau_{0}=120$, gives a \textit{dimensionful\/}
bandwidth $\Delta\overline{\omega} = (120/5\mu s)\Delta\omega = 1.44$ GHz.
This is well within the range of commercially available arbitrary waveform
generators (AWG). From the known values of the control operation time $T$ and
the dimensionless parameters in Table~\ref{table1} it is straightforward to 
determine the values of dimensionful Hamiltonian parameters (see formulas in
Appendix~\ref{AppA}). With these values the improved control $\bldF (t)$ is 
fully specified.

\section{Performance impact of non-ideal control}
\label{sec4}

In this Section we examine the robustness of NOC state preparation to two types
of imperfections in the AWG that generates the control field: 
(i)~finite-precision control parameters; and (ii)~timing jitter. Although our 
interest here is in the effects of these control errors on NOC performance 
gains, we do not mean to imply that these are the only types of errors that 
can afflict a qubit.

\subsection{Finite-precision}
\label{sec4_1}

The NOC formalism requires a nominal control $\bldF_{0}(t)$ that can produce a 
good approximation $U_{0,f}$ to a target unitary $U_{tgt}$. The NOC modification
$\DbfF (t)$ is then optimum relative to $\bldF_{0}(t)$. Altering the nominal 
control field $\bldF_{0}(t)\rightarrow\bldF^{\prime}_{0}(t)$ may cause
$\DbfF (t)$ to become sub-optimal relative to $\bldF^{\prime}_{0}(t)$. Because 
the hardware (viz.~AWG) used to produce $\bldF_{0}(t)$ has limited precision it 
is important to examine the degree of precision the control parameters must have
to achieve the NOC performance gains.

We have seen that the parameter values appearing in Table~\ref{table1} produce 
an approximate Bell state $|\psi^{NOC}_{f}\rangle$ (Eq.~(\ref{NOCBellstate})) 
with error probability $\epsilon_{NOC} = 2.58\times 10^{-6}$. To examine the 
robustness of this result to finite-precision Hamiltonian parameters we shifted 
one parameter away from its optimum value in its fourth significant digit, while
keeping the remaining parameters fixed. This alters the nominal control 
$\bldF_{0}(t)\rightarrow\bldF^{\prime}_{0}(t)$. To study the effect on 
performance we numerically simulated the Schrodinger dynamics driven by 
$\calH_{2}[\bldF^{\prime}(t)]$ using $\bldF^{\prime}(t) = \bldF_{0}^{\prime}
(t) + \DbfF (t)$ with $\DbfF (t)$ the NOC modification relative to 
$\bldF_{0}(t)$. Tables~\ref{table3}--\ref{table9}\hspace{-0.6em}
\begin{table}
\caption{\label{table3}Sensitivity of the state preparation error probability 
$\epsilon$ to small variation of $\eta_{4}$ away from its optimum value
(marked with an asterisk). The other Hamiltonian parameters remain at their
optimum (Table~\ref{table1}) values.}
\begin{center}
\begin{tabular}{|cc|} \hline
$\eta_{4}$ & $\epsilon$\\\hline
$4.525\times 10^{-4}$ & $1.90\times 10^{-5}$ \\
$^{\ast}4.526\times 10^{-4}$ & $2.58\times 10^{-6}$\\
$4.527\times 10^{-4}$ & $1.35\times 10^{-5}$\\\hline
\end{tabular}
\end{center}
\end{table}
\begin{table}
\caption{\label{table4}Sensitivity of the state preparation error probability 
$\epsilon$ to small variation of $\lambda$ away from its optimum value
(marked with an asterisk). The other Hamiltonian parameters remain at their 
optimum (Table~\ref{table1}) values.}
\begin{center}
\begin{tabular}{|cc|} \hline
$\lambda$ & $\epsilon$\\\hline
$9.578$ & $7.89\times 10^{-6}$ \\
$^{\ast}9.579$ & $2.58\times 10^{-6}$\\
$9.580$ & $3.35\times 10^{-5}$\\\hline
\end{tabular}
\end{center}
\end{table}
\begin{table}
\caption{\label{table5}Sensitivity of the state preparation error probability 
$\epsilon$ to small variation of $d_{1}$ away from its optimum value
(marked with an asterisk). The other Hamiltonian parameters remain at their 
optimum (Table~\ref{table1}) values.}
\begin{center}
\begin{tabular}{|cc|} \hline
$d_{1}$ & $\epsilon$\\\hline
$1.385$ & $2.37\times 10^{-5}$ \\
$^{\ast}1.386$ & $2.58\times 10^{-6}$\\
$1.387$ & $1.99\times 10^{-5}$\\\hline
\end{tabular}
\end{center}
\end{table}
\begin{table}
\caption{\label{table6}Sensitivity of the state preparation error probability 
$\epsilon$ to small variation of $d_{2}$ away from its optimum value
(marked with an asterisk). The other Hamiltonian parameters remain at their 
optimum (Table~\ref{table1}) values.}
\begin{center}
\begin{tabular}{|cc|} \hline
$d_{2}$ & $\epsilon$\\\hline
$9.621$ & $5.31\times 10^{-5}$ \\
$^{\ast}9.622$ & $2.58\times 10^{-6}$\\
$9.623$ & $2.70\times 10^{-5}$\\\hline
\end{tabular}
\end{center}
\end{table}
\begin{table}
\caption{\label{table7}Sensitivity of the state preparation error probability 
$\epsilon$ to small variation of $d_{3}$ away from its optimum value
(marked with an asterisk). The other Hamiltonian parameters remain at their 
optimum (Table~\ref{table1}) values.}
\begin{center}
\begin{tabular}{|cc|} \hline
$d_{3}$ & $\epsilon$\\\hline
$8.904$ & $4.68\times 10^{-6}$ \\
$^{\ast}8.905$ & $2.58\times 10^{-6}$\\
$8.906$ & $8.05\times 10^{-6}$\\\hline
\end{tabular}
\end{center}
\end{table}
\begin{table}
\caption{\label{table8}Sensitivity of the state preparation error probability 
$\epsilon$ to small variation of $d_{z}$ away from its optimum value
(marked with an asterisk). The other Hamiltonian parameters remain at their 
optimum (Table~\ref{table1}) values.}
\begin{center}
\begin{tabular}{|cc|} \hline
$d_{z}$ & $\epsilon$\\\hline
$0.917$ & $1.67\times 10^{-4}$ \\
$^{\ast}0.918$ & $2.58\times 10^{-6}$\\
$0.919$ & $8.70\times 10^{-5}$\\\hline
\end{tabular}
\end{center}
\end{table}
\begin{table}
\caption{\label{table9}Sensitivity of the state preparation error probability 
$\epsilon$ to small variation of $d_{xy}$ away from its optimum value
(marked with an asterisk). The other Hamiltonian parameters remain at their 
optimum (Table~\ref{table1}) values.}
\begin{center}
\begin{tabular}{|cc|} \hline
$d_{xy}$ & $\epsilon$\\\hline
$4.330$ & $5.59\times 10^{-4}$ \\
$^{\ast}4.331$ & $2.58\times 10^{-6}$\\
$4.332$ & $1.33\times 10^{-4}$\\\hline
\end{tabular}
\end{center}
\end{table}
show the state preparation error probability $\epsilon$ found as we varied
each Hamiltonian parameter. For example, Table~{\ref{table3} shows that as 
$\eta_{4}$ is varied away from its optimum value by $(+1/-1)$ in its fourth 
significant digit the error probability shifts from $2.58\times 10^{-6}$ at 
optimum to $(1.35/1.90)\times 10^{-5}$. Examination of the Tables shows that 
performance is most sensitive to $d_{xy}$. Note however that if the control 
parameters can be controlled to $1$ part in $10,000$ or better, then the NOC
performance gains are achievable. This corresponds to parameters with 
$14$-bit precision (viz.~$1$ part in $2^{14}=16,384$). On the other hand, 
$13$-bit precision ($1$ part in $2^{13}= 8192$) will cause uncertainty in the 
fourth significant digit and should result in non-optimal NOC performance.

\subsection{Timing jitter}
\label{sec4_2}

Timing jitter arises from timing errors in the clock used in an AWG. Ideally the
clock outputs a sequence of ``ticks'' with constant time separation $T_{clock}$
derived from an oscillation with phase $\phi (t) = 2\pi f_{clock}t$, where 
$f_{clock}=1/T_{clock}$. In a real clock, however, the time $T$ between ticks
is a stochastic process $T = T_{clock}+\delta t$, where the stochastic timing
error $\delta t$ has vanishing time average $\overline{\delta t} = 0$ and
standard deviation $\sigma_{t} = \sqrt{\overline{\delta t^{2}}}$ that
quantifies the spread of tick intervals about $T_{clock}$. The timing error
$\delta t$ causes a phase error $\delta\phi = (2\pi f_{clock})\delta t$ which
has $\overline{\delta\phi} = 0$ and standard deviation $\sigma_{\phi} = 
\sqrt{\overline{\delta\phi^{2}}}$. It is straightforward to show 
\cite{Peng&Gaitan} that $\sigma_{\phi} = (2\pi f_{clock})\sigma_{t}$.

Timing jitter causes phase noise in the TRP nominal control $\bldF_{0}(t)$.
Specifically, the TRP quartic twist profile $\phi_{4}(t) = (\eta_{4}/2\lambda )
\tau^{4}$ picks up phase noise $\delta\phi (\tau )$ due to the timing 
error $\delta\tau$ ( we switch over to dimensionless time): $\phi_{4}(\tau ) 
\rightarrow \phi_{4}(\tau ) +\delta\phi (\tau )$. This causes the TRP
control to twist incorrectly and yields a noisy nominal control $\bldF_{0}
(\tau ) \rightarrow \bldF_{0}^{\prime}(\tau )$. As the phase noise cannot be 
known in advance, it is not possible to determine a control modification that 
is optimal for $\bldF_{0}^{\prime}(\tau )$. All one can do is calculate the
control modification $\DbfF (\tau )$ which is optimal for the jitter-free
control $\bldF_{0}(\tau )$ and add it to the noisy nominal control 
$\bldF^{\prime}_{0}(\tau )$. Since $\DbfF (\tau )$ is not optimal for 
$\bldF^{\prime}_{0}(\tau )$, timing jitter is expected to reduce NOC 
performance.

To quantitatively study the effects of timing jitter on NOC performance we
modelled the phase noise $\delta\phi (\tau )$ as shot noise and used the 
model to generate numerical realizations of the phase noise $\delta\phi
(\tau )$. The details of the model and the protocol used to generate a noise
realization are described in Ref.~\cite{Peng&Gaitan} (see also 
Ref.\cite{noisyQuAdS}). For each noise realization we determined the state 
$|\psi (\tau )\rangle$ by numerically integrating the Schrodinger dynamics 
driven the noisy control $\bldF^{\prime}(\tau ) =\bldF^{\prime}_{0}(\tau ) 
+ \DbfF (\tau )$ and used it to determine the error probability $\epsilon$ for 
the state produced. We generated $10$ phase noise realizations $\delta\phi_{i}
(\tau )$ and determined $10$ error probabilities $\epsilon_{i}$. From these we 
determined the average error probability $\overline{\epsilon}$ and standard 
deviation $\sigma_{\epsilon}$. These were used to estimate the noise-averaged 
performance of NOC state preparation with TRP providing the nominal control. 
For timing jitter $\sigma_{t} = 5.03$ ps and clock frequency $f_{clock} = 1$ GHz
(typical of a commercially available AWG), the simulations found 
$\overline{\epsilon}\pm\sigma_{\epsilon} = (1.64\pm 0.16)\times 10^{-5}$. We 
see that timing jitter does impact NOC performance, though the resulting error 
probability remains quite small at jitter levels typical of present-day AWGs. 
In closing, note that for nominal controls whose good performance is not due to 
controllable quantum intereference effects, timing jitter may have less impact 
on NOC performance than for the TRP nominal control considered here.

\section{Discussion}
\label{sec5}

We have presented a procedure for single-shot high-fidelity quantum state
preparation based on NOC theory. We illustrated the procedure by using it to 
prepare a high-fidelity approximation to the Bell state $|\beta_{01}\rangle$.
The resulting state had error probability $\epsilon_{NOC}\sim 10^{-6}$
($10^{-5}$) for ideal  (non-ideal) control. The excellent fidelity of the
final state provides proof-of-principle of the performance gains possible
using NOC, even in the presence of control imperfections.

We have assumed throughout this paper that the qubit longitudinal ($T_{1}$)
and transverse ($T_{2}$) relaxation times are long compared to the state 
preparation time $T_{sp}$. This assumption is essential for any discussion of 
fault-tolerant quantum computing and error correction as it ensures that the 
qubit state does not decohere away before the error-syndrome extraction
circuit can be applied, and likely errors identified. When $T_{sp} \ll T_{1},
T_{2}$, control imperfections may be anticipated to be the primary source of 
errors during the time for state preparation, and the qubit environment
a secondary source. On the other hand, when $T_{1},T_{2} \approx T_{sp}$, 
the qubits are of sufficiently poor quality that errors from the qubit 
environment can be expected to be (at least) as bad as the types of errors
we have examined in this paper. Our NOC procedure for improving state 
preparation does not remove the need for high-quality qubits as the objects of 
these controls.

In this paper we used TRP to provide a good starting control to be improved
by NOC. Other controls could be used as well. It would be very interesting 
to use GRAPE \cite{GRAPE} to provide the input control for the NOC formalism 
and to examine what kind of performance gains are possible. We plan to carry 
out such a study in future work. 

The high-fidelity NOC preparation of $|\beta_{01}\rangle$ can be 
straightforwardly incorporated into Knill's procedure for fault-tolerant
logical Bell state preparation \cite{knill}. This requires working with two
codeblocks of a [$4$,$2$,$2$] quantum error detecting code. In Ref.~\cite{knill}
each physical Bell state is prepared using a Hadamard and a CNOT gate. With NOC, 
each physical Bell state is prepared in a single-shot as described in 
Section~\ref{sec3}, reducing the depth of state preparation by a factor of two. 
As many Bell states are needed in a large-scale quantum computation, the 
cummulative effect of this reduction could be significant. 

As noted in Section~\ref{sec2}, the NOC state preparation procedure can be used 
to prepare $n$-qubit states. It would be interesting to examine the 
effectiveness of this procedure for single-shot high-fidelity preparation of 
logical states in small to moderate size quantum error correcting codes. 
Efforts along this direction are currently underway.

Finally, we have included as supplementary material the MATLAB source files 
used to obtain the numerical results presented in this paper.

\begin{acknowledgements}
F. G. thanks T. Howell III for continued support.
\end{acknowledgements}

\begin{coi}
{\footnotesize\textbf{Conflicts of Interest}~The authors declare 
that they have no conflict of interest.}
\end{coi}

\appendix

\section{TRP essentials and the dimensionless Hamiltonian 
$\overline{\calH}_{det}^{0}(\tau )$}
\label{AppA}

\textbf{(i) TRP Essentials:}~To introduce twisted rapid passage (TRP) 
\cite{orgTRP,expTRP} we consider a single-qubit interacting with an external 
field $\bldF (t)$ via the Zeeman interaction $H(t) = -\bsig\cdot\bldF (t)$, 
where the $\sigma_{i}$ are the Pauli matrices ($i=x,y,z$). TRP is a 
generalization of adiabatic rapid passage (ARP). In ARP the control field 
$\bldF (t)$ is slowly inverted over a time $T$ with $\bldF (t) = at\hat{\bldz} 
+ b\hat{\bldx}$. In TRP, however, the control field is allowed to twist in the 
$x$-$y$ plane with time-varying azimuthal angle $\phi (t)$, while simultaneously
undergoing inversion along the $z$-axis:
\begin{displaymath}
\bldF_{0}(t) = at\hat{\bldz} + b\cos\phi (t)\hat{\bldx} + b\sin\phi (t)
                                \hat{\bldy}.
\end{displaymath}
Here $t\in [-T/2,T/2]$ and we consider TRP with \textit{nonadiabatic\/} 
inversion. As shown in Ref.~\cite{orgTRP}, the qubit undergoes resonance when
\begin{equation}
at -\frac{\hbar}{2}\frac{d}{dt}\phi (t) = 0.
\label{rescond}
\end{equation}
For polynomial twist, the twist profile $\phi (t)$ takes the form
\begin{displaymath}
\phi_{n}(t) = (2/n)Bt^{n}.
\end{displaymath}
In this case Eq.~(\ref{rescond}) has $n-1$ roots, though only real-valued roots 
correspond to resonance. Ref.~\cite{orgTRP} showed that for $n\geq 3$, the qubit
undergoes resonance \textit{multiple\/} times during a \textit{single\/} TRP
sweep: (i)~for all $n\geq 3$ when $B>0$; and (ii)~for odd $n\geq 3$ when $B<0$.
In this paper we restrict ourselves to $B>0$ and quartic twist ($n=4$). During
quartic twist the qubit passes through resonance at time $t=0,\pm\sqrt{a/
\hbar B}$. It is thus possible to alter the time separating the resonances by 
varying the TRP sweep parameters $B$ and $a$. Ref.~\cite{orgTRP} showed that
these multiple resonances have a strong influence on the qubit transition 
probability, allowing transitions to be strongly enhanced or suppressed through
small variation of the sweep parameters. Ref.~\cite{expTRP} observed these
quantum interference effects in the transition probability using NMR and found
excellent agreement between theory and experiment. Subsequently, TRP controls
were used to produce a high-fidelity universal set of quantum gates
\cite{TRPQC1,TRPQC2,TRPQC3,TRPQC4} which were further improved using NOC
\cite{Peng&Gaitan}.\\

\noindent\textbf{(ii) Dimensionless Hamiltonian:}~The two-qubit Hamiltonian $\calH_{2}
(t)$ describes qubits coupled to an external control field $\bldF (t)$ via the 
Zeeman interaction and to each other via an anisotropic Heisenberg interaction. 
In the lab frame the control field $\bldF_{0}(t)$ is
\begin{equation}
\bldF_{0}(t) = B_{0}\hat{\bldz} + B_{rf}\cos\phi_{rf}(t)\hat{\bldx} - B_{rf}
                              \sin\phi_{rf}(t)\hat{\bldy},
\label{TRPconfld}
\end{equation}
and $\calH_{2}(t)$ is ($\hbar = 1$)
\begin{displaymath}
\calH^{0}_{2}(t) = -\sum_{i=1}^{2}\frac{\gamma_{i}}{2}\bsig\cdot\bldF_{0}(t) +
            \frac{\pi}{2}\left[ J_{z}\sz^{1}\sz^{2} + J_{xy}\left( \sx^{1}\sx^{2}
              +\sy^{1}\sy^{2}\right)\right] .
\end{displaymath}
Introducing $\omega_{i} = \gamma_{i}B_{0}$ and $\omega_{i}^{rf} = \gamma_{i}
B_{rf}$ ($i=1,2$) gives
\begin{eqnarray}
\calH^{0}_{2}(t) & = & -\frac{\omega_{1}}{2}\sz^{1} -\frac{\omega_{1}^{rf}}{2}
                              \left[\cos\phi_{rf}\sx^{1} - \sim\phi_{rf}
                                    \sy^{1}\right] \nonumber \\
 & & \hspace{0.5in} -\frac{\omega_{2}}{2}\sz^{2} -\frac{\omega_{2}^{rf}}{2}
            \left[\cos\phi_{rf}\sx^{2} -\sin\phi_{rf}\sy^{2}\right] \nonumber \\
 & &\hspace{1.0in} -\frac{\pi}{2}\left[ J_{z}\sz^{1}\sz^{2} + J_{xy}\left(
         \sx^{1}\sx^{2} + \sy^{1}\sy^{2}\right)\right] .
\label{Hamtmp1}
\end{eqnarray}
We switch to the detector frame \cite{expTRP,quadTRP} by applying the unitary
\begin{equation}
U_{det}(t) = \exp\left[\frac{i}{2}\phi_{det}(t)\left(\sz^{1}+\sz^{2}\right)
                             \right] .
\label{Udet}
\end{equation}
The detector frame Hamiltonian is
\begin{equation}
\calH_{det}^{0}(t) = U^{\dagger}_{det}(t)\calH^{0}_{2}(t)U_{det}(t) 
                  -iU^{\dagger}_{det}(t)\frac{d}{dt}U_{det}(t)  . 
\label{detHam_v0}
\end{equation}
Inserting Eqs.~(\ref{Hamtmp1}) and (\ref{Udet}) into (\ref{detHam_v0}) gives
\begin{eqnarray}
\calH^{0}_{det} & = &  -\frac{\left(\omega_{1}+\dot{\phi}_{det}\right)}{2}
                                      \sz^{1} -\frac{\omega_{1}^{rf}}{2}\left[
               \cos\left(\phi_{det}-\phi_{rf}\right)\sx^{1}+\sin\left(\phi_{det}-\phi_{rf}
                  \right)\sy^{1}\right] \nonumber\\
 & & \hspace{0.3in} -\frac{(\omega_{2}+\dot{\phi}_{det})}{2}\sz^{2}
           -\frac{\omega_{2}^{rf}}{2}\left[\cos\left(\phi_{det}-\phi_{rf}\right)
             \sx^{2} + \sin\left(\phi_{det}-\phi_{rf}\right)\sy^{2}\right] 
                    \nonumber\\
& & \hspace{0.75in} -\frac{\pi}{2}\left[ J_{z}\sz^{1}\sz^{2} + J_{xy}\left( 
            \sx^{1}\sx^{2} +\sy^{1}\sy^{2}\right)\right] .
\label{detHam_v1}
\end{eqnarray}

To produce a TRP sweep in the detector frame it is necessary to sweep 
$\dot{\phi}_{det}$ and $\dot{\phi}_{rf}$ through a Larmor resonance frequency
\cite{TRPQC2}. Without lose of generality we chose to sweep through the Larmor
frequency $\omega_{2}$ and introduce a detuning $\Delta$:
\begin{eqnarray}
\dot{\phi}_{det}(t) & = & \omega_{2} + \frac{2at}{\hbar} + \Delta \nonumber\\
 \phi_{rf}(t) & = & \phi_{det}(t) -\phi_{trp}(t) .
\label{phidefs}
\end{eqnarray}
Here $a$ is the TRP inversion rate and for quartic twist $\phi_{trp}(t) =
\phi_{4}(t) = (1/2)Bt^{4}$. Introducing $\delta\omega = \omega_{1}-
\omega_{2}$, Eq.~(\ref{detHam_v1}) becomes
\begin{eqnarray}
\calH_{det}^{0}(\tau ) & = & \left[ -\frac{(\delta\omega + \Delta )}{2}
            +\frac{at}{\hbar}\right]\sz^{1} -\frac{\omega_{1}^{rf}}{2}\left[
               \cos\phi_{4}(\tau )\sx^{1} + \sin\phi_{4}(\tau )\sy^{1}\right]
                  \nonumber\\
 & & \hspace{0.5in}\left[ -\frac{\Delta}{2}+\frac{at}{\hbar}\right]\sz^{2}
      -\frac{\omega_{2}^{rf}}{2}\left[\cos\phi_{4}(\tau )\sx^{2}+
           \sin\phi_{4}(\tau )\sy^{2}\right] \nonumber\\
 & & \hspace{1.0in} -\frac{\pi}{2}\left[ J_{z}\sz^{1}\sz^{2} +J_{xy}
               \left(\sx^{1}\sx^{2} + \sy^{1}\sy^{2}\right)\right] .
\label{detHam_v2}
\end{eqnarray}
It proves convenient for the numerical simulations to transform 
$\calH^{0}_{det}(t)$ to dimensionless form. To that end we introduce 
$b_{i} = \hbar\omega_{i}^{rf}/2$ ($i=1,2$), $\lambda =\hbar a/b^{2}_{2}$, the
dimensionless time $\tau = (a/b_{2})t$, and the dimensionless inversion time
$\tau_{0} = (a/b_{2})T$. The dimensionless quartic twist profile is $\phi_{4}
(\tau ) = (\eta_{4}/2\lambda )\tau^{4}$, where $\eta_{4} = \hbar Bb_{2}^{2}
/a^{3}$. Multiplying both sides of Eq.~(\ref{detHam_v2}) by $b_{2}/a$ gives 
the dimensionless Hamiltonian
\begin{eqnarray}
\overline{\calH}_{det}^{0} & = & \left[ -\frac{(d_{1}+d_{2})}{2} + 
    \frac{\tau}{\lambda}\right]\sz^{1} -\frac{d_{3}}{\lambda}\left[
           \cos\phi_{4}(\tau )\sx^{1} +\sin\phi_{4}(\tau )\sy^{1}\right]
           \nonumber\\
 & & \hspace{0.5in} \left[ -\frac{d_{2}}{2} + \frac{\tau}{\lambda}\right]
          \sz^{2} -\frac{1}{\lambda}\left[\cos\phi_{4}(\tau )\sx^{2} +
           \sin\phi_{4}(\tau )\sy^{2}\right] \nonumber\\
 & & \hspace{1.0in} -\frac{\pi}{2}\left[ d_{z}\sz^{1}\sz^{2} + d_{xy}\left(
          \sx^{1}\sx^{2} + \sy^{1}\sy^{2}\right)\right] ,
\label{AppdimHam}
\end{eqnarray}
where: (i)~$d_{1} = (b_{2}/a)\delta\omega$; $d_{2} = (b_{2}/a)\Delta $; 
$d_{3} = b_{1}/b_{2}$; $d_{z} = (b_{2}/a)J_{z}$; and $d_{xy} =(b_{2}/a)J_{xy}$.
This is the dimensionless Hamiltonian that appears in 
Eqs.~(\ref{dimHam})--(\ref{intHam}). We see that $\overline{\calH}_{det}^{0}
(\tau )$ depends on the TRP sweep parameters ($\lambda$, $\eta_{4}$) as well as
the coupling parameters ($d_{1}$, $d_{2}$, $d_{3}$, $d_{z}$, $d_{xy}$).
The latter set of parameters are dimensionless versions of, respectively, 
the Larmor frequency difference $\delta\omega$, the detuning $\Delta$, the
Zeeman coupling ration $b_{1}/b_{2}$, and the Heisenberg couplings $J_{z}$ 
and $J_{xy}$.

\end{document}